\documentclass[aps,prb,twocolumn,showpacs]{revtex4}

\usepackage{graphicx}
\usepackage{amsmath}
\usepackage{times}
\usepackage{bm}
\usepackage{amssymb}

\begin{document}
\title{Theory of Edge States in Systems with Rashba Spin-Orbit Coupling}
\author{A. Reynoso}
\affiliation{Instituto Balseiro and Centro At\'omico Bariloche, Comisi\'on
Nacional de Energ\'\i a At\'omica, 8400 San Carlos de Bariloche, Argentina.}
\author{Gonzalo Usaj}
\affiliation{Instituto Balseiro and Centro At\'omico Bariloche, Comisi\'on
Nacional de Energ\'\i a At\'omica, 8400 San Carlos de Bariloche, Argentina.}
\author{M. J. S\'anchez}
\affiliation{Instituto Balseiro and Centro At\'omico Bariloche, Comisi\'on
Nacional de Energ\'\i a At\'omica, 8400 San Carlos de Bariloche, Argentina.}
\author{C. A. Balseiro}
\affiliation{Instituto Balseiro and Centro At\'omico Bariloche, Comisi\'on
Nacional de Energ\'\i a At\'omica, 8400 San Carlos de Bariloche, Argentina.}
\date{\today}

\begin{abstract}
We study the edge states in a two dimensional electron gas with a transverse
magnetic field and Rashba spin-orbit coupling. In the bulk, the interplay
between the external field perpendicular to the gas plane and the spin-orbit
coupling leads to two branches of states that, within the same energy
window, have different cyclotron radii. For the edge states, surface
reflection generates hybrid states with the two cyclotron radii. We analyze
the spectrum and spin structure of these states and present a semiclassical
picture of them.
\end{abstract}
\maketitle

\section{Introduction}
Since the seminal spin-transistor proposal by Datta and Das,\cite{DattaD90}
it has been recognized that the spin-orbit interaction may be a useful tool
to manipulate and control the spin degree of freedom of the charge carriers.
This opens novel opportunities for the developing field of spintronics.\cite
{Spintronicsbook} The challenging task of building spin devices based purely
on semiconducting technology requires to inject, control and detect spin
polarized currents without using strong magnetic fields. For this purpose,
the spin-orbit coupling may be a useful intrinsic effect that links
currents, spins and external fields. During the last years a number of
theoretical and experimental papers were devoted to study the effect of
spin-orbit coupling on the electronic and magnetotransport properties of two
dimensional electron gases (2DEG).\cite{AronovG93,Hirsch99,MolenkampSB01,ZumnuhlMMCG02,GovernaleZ02,StredaS03,MishchenkoH03,SchliemannL03,SinovaCSJM03,ChenHPDG03,Rashba04}
This is endorsed by the fact that, in some of the semiconducting
heterostructures used to confine the electron or hole gas, the spin orbit
interaction is large. Moreover it may be varied by changing carrier
densities or gating with external electric fields.\cite
{NittaATE97,MillerZMLGCG03}

In many transport experiments in 2DEG with a transverse magnetic field,
including quantum Hall effect\cite{Halperin82} and transverse magnetic
focusing,\cite{vanHouten89,BeenakkerH91,PotokFMU02} edge states play a
central role. To our knowledge, a detailed analysis of the effect of the
spin-orbit coupling on the edge states has not been done yet. In this paper
we present a theory for edge states in 2DEG with transverse magnetic fields
and a Rashba term describing the spin orbit interaction.\cite{Rashba60}

First we focus on the quantum mechanical solution. By numerical
diagonalization of the Hamiltonian in a truncated Hilbert space we calculate
the energy spectrum and the wavefunctions that, as we show below, present an
intricate structure. Then we resort to a semiclassical analysis to interpret
and illustrate the nature of the edge states in the high energy or low field
limit.

Our starting point is a 2DEG with Rashba coupling and an external magnetic
field $B$ perpendicular to the plane containing the electron gas: 
\begin{equation}
H\! = \! \frac{1}{2m^{*}}(P_{x}^{2}\! + \! P_{y}^{2})\!+\frac{\alpha }{\hbar 
}(P_{y}\mathbf{\sigma }_{x}\! - \! P_{x}\mathbf{\sigma }_{y})\!+\frac{1}{2}%
g\mu _{B}B\mathbf{\sigma }_{z}\! + \! V(x)  \label{e1}
\end{equation}
where $m^{*}$ is the effective mass of the carriers, $P_{\eta }\! = \!
p_{\eta }+(e/c)A_{\eta }$, with $p_{\eta }$ and $A_{\eta }$ being the $\eta $%
-component of the momentum and vector potential respectively, $\alpha $ is
the Rashba coupling parameter, $\mathbf{\sigma }_{\eta }$ are the Pauli
matrices and $g$ is the gyromagnetic factor. The last term $V(x)$ is the
lateral confining potential. For simplicity, from hereon we consider a hard
wall potential that confines the electrons in the transverse $x$-direction: $%
V(x)\! = \! 0$ for $0\leq x\leq L$ and infinite otherwise.

\section{The Quantum Solution}

In the geometry where electrons are confined in the $x$-direction, it is
convenient to use the Landau gauge $\mathbf{A}\! = \! (0,xB,0)$ and write
the wavefunction in the form: 
\begin{equation}
\mathbf{\Psi }(x,y)\! = \! e^{\mathrm{i} ky}\mathbf{\varphi }(x) \; ,
\end{equation}
with the function $\mathbf{\varphi }(x)$ expanded in the basis set of the
infinite potential well 
\begin{equation}
\mathbf{\varphi }(x)\! = \! \sqrt{\frac{2}{L}} \sum_{n}\sin \left(\frac{\pi n%
}{L}x\right)\binom{a_{n}}{b_{n}} \,.
\end{equation}

The Schr\"{o}dinger equation $H\mathbf{\Psi }\!=\!E\mathbf{\Psi }$ leads to
the following equations for the spinors 
\begin{eqnarray}
&&\left( \frac{\hbar ^{2}}{2m^{\ast }}\left( \frac{\pi l}{L}\right) ^{2}+%
\frac{g}{2}\mu _{B}B\mathbf{\sigma }_{z}-E\right) \left( 
\begin{array}{c}
a_{l} \\ 
b_{l}
\end{array}
\right) \!=\!  \notag  \label{equ} \\
&&\sum_{m}[(F_{lm}\!-\!G_{lm})\mathbf{\sigma }^{-}\!-\!(F_{lm}+G_{lm})%
\mathbf{\sigma }^{+}\!-\!M_{lm}]\left( 
\begin{array}{c}
a_{m} \\ 
b_{m}
\end{array}
\right)   \notag \\
&&
\end{eqnarray}
with $M_{lm}$, $F_{lm}$ and $G_{lm}$ proportional to the matrix elements of
the operators $(x-x_{0})^{2}$, $(x-x_{0})$ and $\partial /\partial x$
respectively, 
\begin{eqnarray}
M_{lm} &\!=\!&\frac{m^{\ast }\omega _{c}^{2}}{L}\int_{0}^{L}\!\!dx\sin
\left( \frac{\pi l}{L}x\right) (x\!-\!x_{0})^{2}\sin \left( \frac{\pi m}{L}%
x\right)   \notag \\
F_{lm} &\!=\!&\frac{2\alpha }{L}\frac{eB}{\hbar c}\int_{0}^{L}\!\!dx\sin
\left( \frac{\pi l}{L}x\right) (x\!\!-\!\!x_{0})\sin \left( \frac{\pi m}{L}%
x\right)   \notag \\
G_{lm} &\!=\!&\frac{2\alpha }{L}\int_{0}^{L}dx\sin \left( \frac{\pi l}{L}%
x\right) {\frac{\partial }{\partial x}}\sin \left( \frac{\pi m}{L}x\right) .
\end{eqnarray}
Here, $\omega _{c}\!=\!e\left| B\right| /m^{\ast }c$ is the cyclotron
frequency, $\mathbf{\sigma }^{\pm }\!=\!(\mathbf{\sigma }_{x}\pm i\mathbf{%
\sigma }_{x})/2$ and $x_{0}\!=\!-\hbar kc/eB$. We solve these equations in a
truncated Hilbert space disregarding the highest energy states. Typically we
take a matrix Hamiltonian of dimension of a few hundreds and keep the first
thirty states. In all cases the width of the sample $L$ is taken large
enough to have the cyclotron radius $r_{c}$ smaller than $L/2$. The right
and left edge states are then well separated in real space. 
\begin{figure}[t]
\includegraphics[height=8cm,clip]{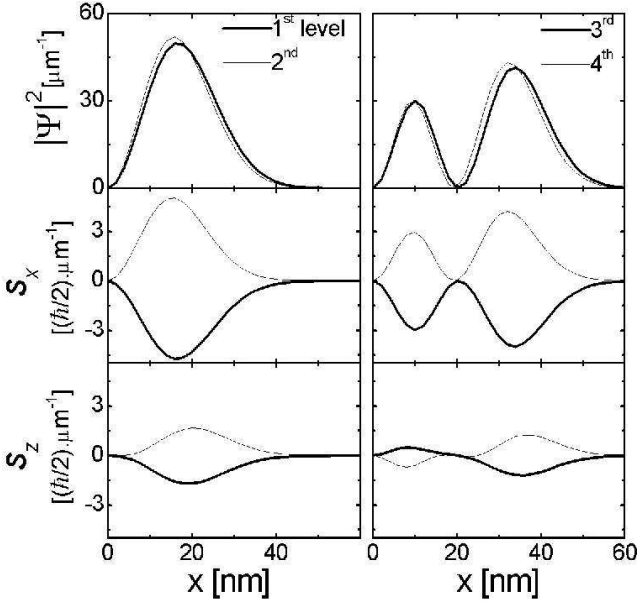}
\caption{Probability and spin densities for the $1^{st}$ and $2^{nd}$ levels
(left panel) and $3^{rd}$ and $4^{th}$ levels (right panel) respectively for 
$x_{0}\!=\!0$, $m^{\ast }\!=\!0.068m_{e}$, $\protect\alpha \!=\!10meVnm$, $%
L\!=\!600nm$ and $B\!=\!2.5T$.}
\label{be}
\end{figure}
For $x_{0}\simeq L/2$ the states are equal to the bulk states, except for
exponential corrections. The wave functions and the energy spectrum
reproduces the known results: in the bulk the spin-orbit coupling mixes the
two spin components and there are two branches of states with energies given
by\cite{Rashba60,BychkovR84} 
\begin{equation}
E_{n}^{\pm }\!=\!\hbar \omega _{c}n\mp \sqrt{E_{0}^{2}+\left( \frac{\alpha }{%
l_{c}}\right) ^{2}2n}  \label{ener}
\end{equation}
with $n\geq 1$ and a single state ($n\!=\!0$) with energy $E_{0}\!=\!\hbar
\omega _{c}/2-g\mu _{B}B/2$. The corresponding eigenfunctions for $n\geq 1$
are 
\begin{equation}
\Psi _{n,k}^{+}(x,y)\!=\!\frac{1}{\sqrt{A_{n}L_{y}}}e^{\mathrm{i}ky}\left( 
\begin{array}{c}
\phi _{n-1}(x) \\ 
-D_{n}\phi _{n}(x)
\end{array}
\right) 
\end{equation}
and 
\begin{equation}
\Psi _{n,k}^{-}(x,y)\!=\!\frac{1}{\sqrt{A_{n}L_{y}}}e^{\mathrm{i}ky}\left( 
\begin{array}{c}
D_{n}\phi _{n-1}(x) \\ 
{\phi _{n}(x)}
\end{array}
\right) \;.
\end{equation}
Here $L_{y}$ is the length of the sample in the $y$-direction, $\phi _{n}(x)$
is the harmonic oscillator wavefunction centered at the coordinate $x_{0}$, $%
A_{n}\!=\!1+D_{n}^{2}$ and 
\begin{equation}
D_{n}\!=\!\frac{\left( \frac{\alpha }{l_{c}}\right) \sqrt{2n}}{E_{0}+\sqrt{%
E_{0}^{2}+\left( \frac{\alpha }{l_{c}}\right) ^{2}2n}}\;.  \label{dn}
\end{equation}
The wavefunction of the state with $n\!=\!0$ is 
\begin{equation}
\Psi _{0,k}(x,y)\!=\!\frac{1}{\sqrt{L_{y}}}e^{\mathrm{i}ky}\left( 
\begin{array}{c}
0 \\ 
{\phi _{0}(x)}
\end{array}
\right) \;.
\end{equation}
\begin{figure}[t]
\includegraphics[height=6.cm,clip]{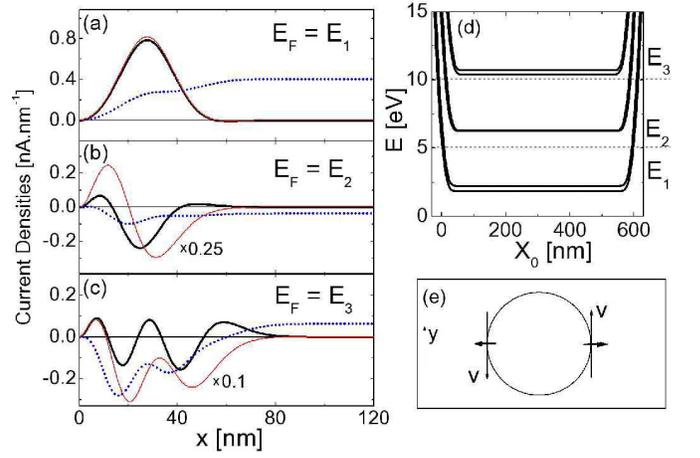}
\caption{Charge and spin density current for three values of the Fermi
energy (left panels). The $s_{z}$ (thick line) and $s_{x}$(dotted line)
current densities are measured in units of $\hbar /2e$. Charge current
density is indicated with a thin line. Note than in (b) and (c) the current
density is multiplied by a numerical factor  indicated in the figure.
In (d) the energy of the first levels versus $x_{0}$ are shown. The three values of the Fermi energy used in (a),
(b) and (c) are indicated. In (e) the semiclassical orbit is shown with  the
velocity and spin direction indicated by arrows. Parameters as in figure 1.}
\label{corr}
\end{figure}
In the bulk ($x_{0}\approx L/2$), the ground state has the spin along the $z$%
-direction. In the excited states the spin is tilted with an expectation
value of its $z$-component $\left\langle \sigma _{z}\right\rangle =\pm
(1-\left| D_{n}\right| ^{2})/A_{n}$ that decreases as $\alpha $ and $n$
increase. The condition $E_{0}^{2}<<2\left({\alpha }/l_{c}\right) ^{2}$, 
that is equivalent to $\left\langle \sigma _{z}\right\rangle \approx 0$
for all $n>0$, is referred as the weak field condition.\cite{BychkovR84} For
large enough $n$, when $E_{0}^{2}<<2 n \left({\alpha }/l_{c}\right) ^{2}$, 
the spin-orbit dominates and $\left\langle \sigma _{z}\right\rangle
\approx 0$.

For high fields or low electron density, the physical properties of the
system are dominated by the states with low quantum number $n$. We first
consider this case and present the results for the first few Landau levels. 

As the momentum $k$ parallel to the edge varies, the center of gravity $x_{0}
$ of the wavefunctions changes and as it approaches the sample edge,the  effect
of the confining potential becomes important generating the $k$-dependent
dispersion of the energy levels.\cite{Halperin82}The interplay of the
spin-orbit coupling and the confining potential produce a tilting of the
spin for \textit{all the edge states}. Edge states probability densities $%
\left| \Psi \right| ^{2}$ and the corresponding spin densities $%
s_{i}\!=\!(\hbar /2)\Psi ^{\dagger }\sigma _{i}\Psi $ are shown in Fig.\ref{be}
for $x_{0}\!=\!0$ . The spin is predominantly in the $xy$-plane even for the
lowest energy edge states and the sign of the spin densities alternates as
the energy increases. The current carried by these states is then polarized
and for the parameters of the figure the polarization is determined by the
Rashba coupling. We use the charge and spin currents operators defined as: $%
J_{y}^{e}\!=\!-e\dot{y}$ for the charge current and $J_{y}^{s_{x}}\!=\!\hbar 
\dot{y}\sigma _{x}/2$ and $J_{y}^{s_{z}}\!=\!\hbar (\dot{y}\sigma
_{z}+\sigma _{z}\dot{y})/4$ for the spin currents.\cite{Rashba03} In these
expressions the velocity operator in the $y$-direction is given by 
\begin{eqnarray}
\dot{y} &\!=\!&\frac{1}{m^{\ast }}\left( \hbar k+\frac{eBx}{c}\right) +\frac{%
\alpha }{\hbar }\sigma _{x}  \notag \\
&\!=\!&(x-x_{0})\frac{eB}{m^{\ast }c}+\frac{\alpha }{\hbar }\sigma _{x}\;.
\end{eqnarray}
The total current densities, defined as $j_{y}^{\nu }(x)\!=\!\sum_{occ}\Psi
_{n}^{\pm \dagger }(x,y)J_{y}^{\nu }\Psi _{n}^{\pm }(x,y)$ where the sum
runs over all occupied states, are shown in Fig.\ref{corr} for different
values of the Fermi energy. The charge and $j_{y}^{s_{z}}$ current densities
are confined at the sample edge indicating that they are due to the edge
states. Conversely the $j_{y}^{s_{x}}$ current density has a non-zero value
inside the sample. The origin of this current can be understood in terms of
the simple semiclassical picture shown in Fig.\ref{corr}e: electrons moving
in the positive (negative) $y$-direction have a positive (negative)
projection of the spin along the $x$-axis. Since the spin is not conserved,
these currents do not necessarily produce spin accumulation in samples with
constrictions or edges perpendicular to the current direction.\cite{Rashba04}

\begin{figure}[t]
\includegraphics[height=6.5cm,clip]{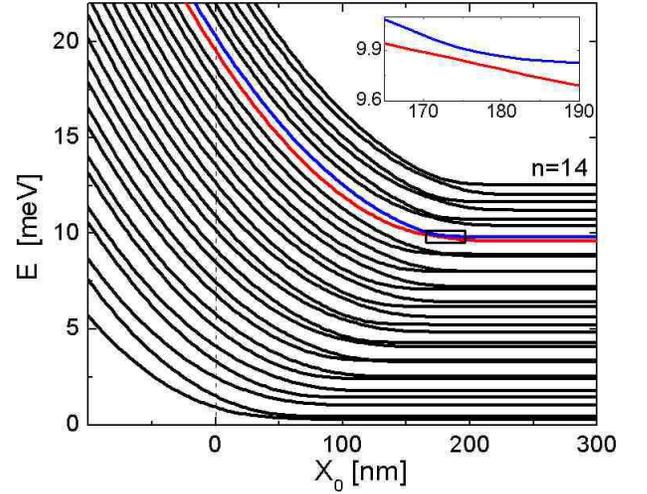}
\caption{Energy spectrum versus center coordinate $x_{0}$ for a system with $%
m^{\ast }\!=\!0.068m_{e}$, $\protect\alpha \!=\!10meVnm$, $L\!=\!600nm$ and $%
B\!=\!0.5T$. Inset: detail of the anticrossing of levels 23 and 24.}
\label{es}
\end{figure}

Let us now consider the low field case where many Landau levels are below
the Fermi energy. The energy spectrum as a function of $x_{0}$ is presented
in Fig.\ref{es}. For the bulk states, the typical energy splitting of the
two branches $(+)$ and $(-)$ is different (see Eq.(\ref{ener})), leading to
a beat in the total energy spectrum. Within the same energy interval, the
two branches have different quantum number $n$ and consequently different
cyclotron radius $r_{c}$. We take\cite{Ferrybook} 
\begin{equation}
r_{c}^{2}\!=\!2\left\langle \Psi _{n}^{\pm }|(x-x_{0})^{2}|\Psi _{n}^{\pm
}\right\rangle
\end{equation}
that for large $n$ gives $r_{c}^{2}\simeq $ $2n(\hbar /m^{\ast }\omega _{c})$%
. According to equation (\ref{ener}), in this limit states with
approximately the same energy belonging to different branches have cyclotron
radius differing in $\Delta r_{c}\simeq 2\alpha /\hbar \omega _{c}$. Note
that the radius difference in this large $n$ approximation does not depend
on $n$. 

For $x_{0}\lesssim r_{c}$ the effect of the confining potential becomes
important and the two bulk branches mix leading to edge states that combine
the two cyclotron radii. This mixing is apparent from the energy spectrum
that presents level anticrossings as shown in the inset of Fig.\ref{es}.

\begin{figure}[t]
\includegraphics[height=7.3cm,clip]{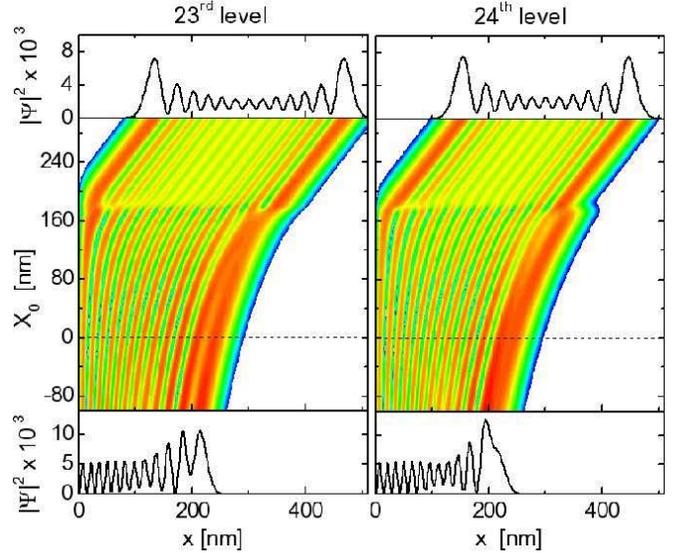}
\caption{Evolution of the wavefunctions of levels 23 and 24 with the center
coordinate $x_{0}$. Parameters as in Figure 1. Upper panel: probability
densities of the levels 23 and 24 for $x_{0}\!=\!300nm$. Central panel:
density plot of the probability densities of the two levels versus $x_{0}$.
Lower panel: probability densities for $x_{0}\!=\!-100nm$.}
\label{modos}
\end{figure}

The behavior of the levels 23 and 24 is illustrated in Fig.\ref{modos}. In
the top panel the figure their probability densities for $x_{0}\! = \! L/2$
are shown. These states correspond to a state of the $(+)$ branch with $n\!
= \! 13$ and a state of the $(-)$ branch with $n\! = \! 10$ respectively.
The $(+)$ branch state radius is larger than the $(-)$ branch one as can be
inferred from the figure. We can follow the evolution of these states as $%
x_{0}$ changes from $x_{0}\! = \! L/2$ to a negative value. A contour plot
illustrating this evolution in shown in the central panel of Fig.\ref{modos}%
. For $x_{0}\sim r_{c}$ a sudden change in the wave function spatial
extension is observed. States belonging to the bulk $(+)$ branch shrink for $%
x_{0}\sim r_{c}$ due to the mixing with the $(-)$ branch states. Conversely,
states belonging to the bulk $(-)$ branch expand for $x_{0}\sim r_{c}$. This
sudden change in the wave function extension is of the order of $\Delta
r_{c} $. The lower panel of Fig.\ref{modos} shows the probability densities
of the two levels for a negative value of $x_{0}$.

\begin{figure}[t]
\includegraphics[height=6cm,clip]{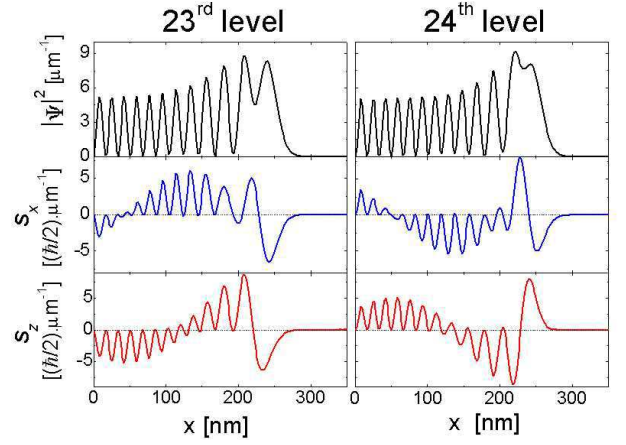}
\caption{Probability densities and the corresponding spin densities for
levels 23 and 24 and $x_{0}\!=\!0$. Parameters as in Fig.\ref{es}.}
\label{spin}
\end{figure}

For $\alpha \!=\!0$ the number of nodes of the wavefunction of a given level
is conserved as $x_{0}$ changes. With spin-orbit coupling, the anticrossing
of energy levels is an indication that the wavefunctions change in character
as $x_{0}$ changes. Then, for a given energy level the number of maxima of
the probability density is no longer conserved as it is shown in Fig.\ref
{modos}. It is also interesting to analyze the spin densities associated to
these states. 
The spin structure of the edge states for $x_{0}\!=\!0$ is shown in Fig.\ref
{spin}. The spin densities $s_{x}$ and $s_{z}$ show an intricate behavior
due to the beating of two contributions. This is a consequence of the mixing
of states that are $(+)$ and $(-)$ in character. As we discuss below the
semiclassical analysis also unveils that edge states are formed by combining
states with different radii and different spin projections.

\section{The Semiclassical Solution}

\subsection{Bulk States}

In a recent approach,\cite{PletyukhovAMB02} both the orbital and the spin
degrees of freedom have been treated semiclassicaly in an extended phase
space. The spin coherent state is defined as 
\begin{equation}
|z\rangle \! = \! \frac{e^{z{\hbar \mathbf{\sigma }}^{+}}}{\sqrt{1+|z|^{2}}}%
|\downarrow \rangle
\end{equation}
where $z$ is a $c$-number. A unit vector associated with the classical spin
is defined in terms of the coherent state $|z\rangle $, which for spin $%
\frac{1}{2}$ reads 
\begin{equation}
\mathbf{n}\! = \! \left\langle z| \mathbf{\sigma }|z\right\rangle \; ,
\end{equation}
with components determined by $z$%
\begin{equation}
n_{1}+in_{2}\! = \! \frac{ 2 z^{*}}{1+|z|^{2}}
\end{equation}
and $n_{1}^{2}+n_{2}^{2}+n_{3}^{2}\! = \! 1$.

The classical phase-space symbol $\mathcal{H} (\mathbf{{\ q, p, n})}$ of the
Hamiltonian defined in Eq.(\ref{e1}) is\cite{PletyukhovAMB02} 
\begin{equation}
\mathcal{H}(\mathbf{q},\mathbf{p},\mathbf{n}) \! = \! \mathcal{H}_{0}(%
\mathbf{q},\mathbf{p})+ \; \frac{\hbar }{2}\mathbf{n}\cdot \mathbf{C(q,p)}
\end{equation}
where the first term is the classical Hamiltonian without spin orbit
coupling and 
\begin{equation}
\mathbf{C(q,p) }\! = \! (\frac{2\alpha }{\hbar ^{2}}(p_{y} + {\frac{e}{c}}
Bx),- \frac{2\alpha }{\hbar ^{2}}p_{x},0\mathbf{)} \, .
\end{equation}

The equations of motion are: 
\begin{equation}  \label{em}
\mathbf{\dot{q}}\! = \! \frac{\partial H}{\partial \mathbf{p}}, \; \mathbf{%
\dot{p}}\! = \! -\frac{\partial H}{\partial \mathbf{q}}, \; \mathbf{\dot{n}}%
\! = \! \mathbf{C}\times \mathbf{n} \;
\end{equation}

whose solutions represent the classical orbits in the extended phase space.
In bulk there is an additional constant of motion besides the energy and the
system is classically integrable. We are interested in the periodic
solutions of Eq.(\ref{em}) from which the action integral can be computed in
order to apply an EBK quantization scheme.\cite{GutzBook} We propose $%
\mathbf{q}\!=\!r(\cos {\omega }t,\sin {\omega }t)$ and replace it in Eq.(\ref
{em}). For $t\!=\!0$ the initial conditions $p_{x}(0)\!=\!p_{y}(0)\!=\!0$
are impossed leading to

\begin{eqnarray}
p_{x} (t) & \! = \! & - \frac{e B}{c} r \; \sin ( {\omega} t ), \; \quad
p_{y} (t) \! = \! 0\, ;  \notag \\
n_{1} (t) & \! = \! & {\frac{\hbar r}{\alpha}} \Delta \cos( \omega t), \;
\,\,\,\,\quad n_{2} (t) \! = \! {\frac{\hbar r}{ \alpha}} \Delta \sin (
\omega t) \, ;  \notag \\
n_{3} (t) & \! = \! & - {\frac{{\hbar}^3}{ 2 {\alpha}^2}} \omega \frac{%
\Delta c }{e B} \! = \! const. \;  \label{po}
\end{eqnarray}
where we have defined $\Delta\! = \! \omega - \omega_c$. The normalization
condition for the spin components gives an algebraic equation of order
fourth in $\Delta$ that to leading order in $\hbar$ (semiclassical limit)
gives 
\begin{equation}  \label{delta}
\Delta \! = \! \pm \frac{\alpha}{ \hbar r} \; .
\end{equation}

Given the two frequencies $\omega = \omega_{c} \pm {\alpha} / {\hbar r}$ one
must go one step further to find an explicit expression for the cyclotron
radius. Replacing Eq.(\ref{po}) evaluated a $t\! = \! 0$ in the equation for
the energy conservation $\mathcal{H} \! = \! E$, and taking into account the
value of $\Delta$ obtained in (\ref{delta}), we find 
\begin{equation}  \label{rs}
r_{\pm}\! = \! \sqrt{\left(\frac{\alpha}{\hbar \omega_c} \right) ^2 + \frac{%
2 E}{m^{*} {\omega_c}^2} } \pm \frac{\alpha}{\hbar \omega_c} \; .
\end{equation}

Therefore for a given energy $E$ the periodic solutions result in two orbits
of radii $r_{\pm}$, frequencies $\omega _{\pm} \! = \! \omega_{c} \mp \frac{%
\alpha}{\hbar r_{\pm}}$ and opposite values of the spin respectively. The
cyclotron radii difference is $\Delta r \equiv r_{+} - r_{-} \! = \! 2
\alpha / \hbar \omega_{c}$ in exact correspondence with the quantum
mechanical estimate that we obtained for large $n$. In these two orbits,
with different radii and frequencies, the electron has the same velocity;
that is $r_{+} \omega_{+} \! = \! r_{-} \omega_{-}\! = \! \sqrt{\big(\frac{%
\alpha}{\hbar} \big) ^2 + \frac{2 E}{m^{*}} }$.

Once we know the periodic solutions in extended phase space we need to
compute the action integral $I$. Following Ref.[\onlinecite{PletyukhovAMB02}%
], the action can be expressed as:

\begin{eqnarray}  \label{actb}
I_{\pm} & \! = \! & \int_{0}^{T_{\pm}} \left( \mathbf{p} \mathbf{{\dot{q}}}
+ {\frac{\hbar}{2}} ( n_{1} {\dot{n}}_{2} - n_{2} {\dot{n}}_{1}) \right) dt
\\
&\! = \! & \frac{e \pi}{c} B r_{\pm}^{2} + \frac{h}{2} \; ,  \notag
\end{eqnarray}

with $T_{\pm }\! = \! 2\pi /\omega _{\pm }$. The EBK quantization rule sets 
\begin{equation}
\frac{I}{\hbar }+\gamma \! = \! 2\pi n
\end{equation}

where $n$ is an integer and $\gamma$ is the sum of the phase shifts acquired
at the turning points of the motion \cite{GutzBook}. For the bulk solutions
the turning points are two caustics giving a total phase shift $\gamma \! =
\! - \pi$. Replacing Eq.(\ref{rs}) in Eq.(\ref{actb}) we finally obtain

\begin{equation}
E^{\pm }\!=\!n\hbar \omega _{c}\mp \frac{\alpha }{l_{c}}\sqrt{2n}\;,
\end{equation}
which is the quantum spectrum for the bulk states (neglecting the zero point
energy) \cite{BychkovR84}. With the notation we emphasize that for a given
quantum index $n$, the $E^{+}$ (energy associated to $r_{+}$) is lower than $%
E^{-}$ (energy associated to $r_{-}$). 
\begin{figure}[t]
\includegraphics[height=5cm,clip]{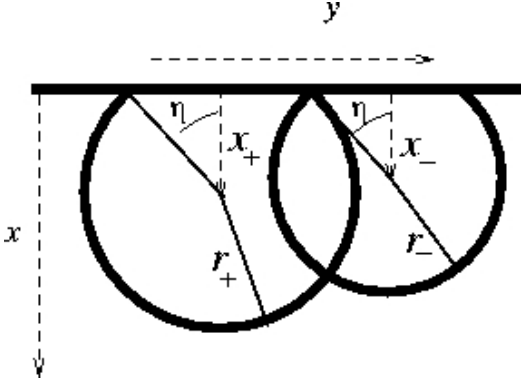}
\caption{Skipping orbit used in the semiclassical calculations drawn up to
the first specular reflection (first period of the $x$ motion). The values
of $x_{+}$ and $x_{-}$ have been enlarged in order to visualize the angle $%
\protect\eta $ (see text for details).}
\label{so}
\end{figure}

\subsection{Edge States}

As we mentioned above, for $x_{0}\sim r_{c}$ the bulk ($+$) and ($-$)
branches mix, and the wavefunctions spatial extension present a remarkable
change. The avoided level crossings structure observed around $x_{0}\sim
r_{c}$ in the energy spectrum is the fingerprint of this behavior. The
semiclassical image that we propose consists of a skipping orbit formed by
a series of translated circular arcs of radii $r_{+}$ and $r_{-}$ and
centers $x_{+}$ and $x_{-}$ in the $x$ direction respectively (see Fig.\ref
{so}). The center coordinate $y$ of these circular arcs changes at each
specular reflection. In Fig.\ref{so} this primitive orbit is plotted for a
complete period of the $x$ motion. The fact that the reflection at the
boundary is specular is guaranteed by the conservation of the modulus of the
velocity ( $r_{+}\omega _{+}\! = \! r_{-}\omega _{-}$) and can be cast in
the form 
\begin{equation}
\cos \eta \! = \! \frac{x_{+}}{r_{+}}\! = \! \frac{x_{-}}{r_{-}}\equiv \zeta
\;,
\end{equation}
being $\eta $ the angle depicted in Fig.\ref{so}.

As it can be inferred from the semiclassical solutions obtained in Eq.(\ref
{po}), to lowest order in $\hbar $ the in plane spin components ($n_{1}$ and 
$n_{2}$) of the orbit $r_{+}$ have opposite signs than those of the orbit $%
r_{-}$ and in both cases is $n_{3}=0$. Therefore the spin conservation is
guaranteed at each specular reflection of the skipping orbit with the
boundary if $x_{+}$ and $x_{-}\sim 0$. Our goal is to perform a
semiclassical quantization employing this classical skipping orbit. The
semiclassical approach is fully justified for angles $\eta \sim \pi /2$ and
we will obtain the energy spectrum and the dispersion relation quite
accurately around $x_{0}\sim 0$. 
\begin{figure}[t]
\includegraphics[height=6.5cm,clip]{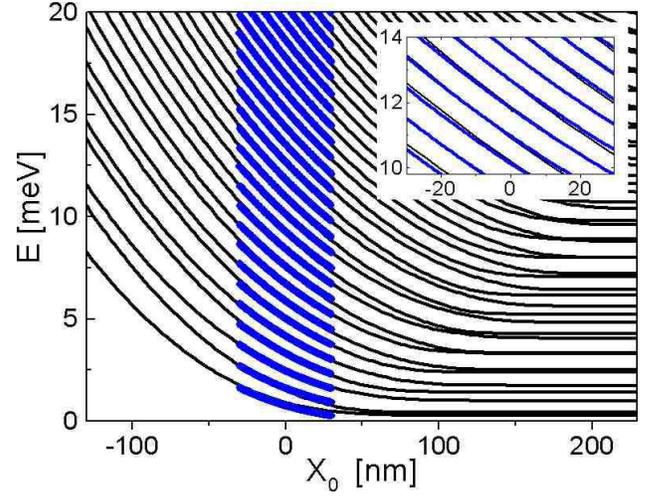}
\caption{Edge states energy spectrum obtained from the semiclassical
approach Eq.(\ref{esemi}) (thick solid lines) together with the exact
quantum results. Around $x_{0}\!=\!0$ both results are almost
indistinguishable even in the low energy region. \ Inset: detail of the
comparisson for $x_{0}\!=\!0$ and intermediate energies. }
\label{rk}
\end{figure}

We proceed analogously to the previous section, taking now into account that
the orbital motion projected on the $x$ axis is periodic with a period $%
T_{s} \equiv T_{1} + T{_2} \! = \! 2 (\pi - \eta)/ \omega_{+} + 2 (\pi -
\eta)/ \omega_{-}$. For the sake of clarity we divide the action integral in
two terms $I \! = \! I_{o} + I_{s}$. The action associated to the orbital
motion is 
\begin{eqnarray}
I_{o} &\! = \!& \int_{-T_{s}/2}^{T_{s}/2} p_{x} \dot{x} dt  \notag \\
&\! = \!&\frac{e}{c}B\omega_{+} r_{+}^{2}\!\int_{-T_{1}/2} ^{T_{1}/2 }
\!\!\sin^{2}(\omega_{+} t) dt  \notag \\
&&+ \frac{e}{c}B \omega_{-} r_{-}^{2}\! \int_{- T_{2}/2} ^{T_{2}/2}\!
\sin^{2}(\omega_{-} t) dt  \notag \\
&\! = \! & \frac {e}{c} B (r_{+}^{2} + r_{-}^{2}) \big(\arccos (-\zeta) +
\zeta \sqrt{ 1 - \zeta^2}) \;.
\end{eqnarray}
For the spin degrees of freedom the action integral is straightforward to
evaluate and gives $I_{s}\! = \! 2 \hbar (\pi -\eta)$. For the skipping
orbit the phase shift is $\gamma \! = \! \pi$, due to the fact that we have
to consider now for each period of motion two bounces with the boundary and
also two caustics. Replacing the obtained values for $I \! = \! I_{o} +
I_{s} $ and $\gamma$ into Eq.(\ref{actb}) we finally obtain 
\begin{equation}  \label{esemi}
E(\zeta) \! = \! \frac{\pi}{2} \frac{ \hbar \omega_c \left( n - \frac{3}{2}
- \frac{1}{\pi} \arccos \zeta \right)} {\arccos (-\zeta) + \zeta \sqrt{ 1 -
\zeta^2}} - m^* \left( \frac{\alpha}{\hbar}\right)^2 \; ,
\end{equation}
as a function of the parameter $\zeta$.

\begin{figure}[t]
\includegraphics[height=6cm,clip]{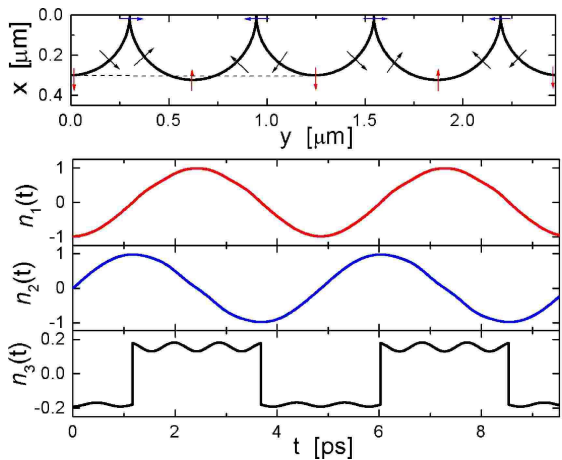}
\caption{The numerically integrated semiclassical results. Semiclassical
skipping orbit with the in-plane spin projection  indicated by arrows
(upper panel). The three components of $\mathbf{n}$ as a function of time
(lower panels). The energy of the orbit is about $37\hbar \protect\omega _{c}
$. Parameters as in Fig. 3.}
\label{sce}
\end{figure}

For $x_{+}\! = \! x_{-}\! = \! 0$, is $\zeta \! = \! 0$ and the energy
levels are 
\begin{equation}
E(0)\! = \! n\hbar \omega _{c}-m^{\ast }\left( \frac{\alpha }{\hbar }\right)
^{2}\;.  \label{e0}
\end{equation}
To obtain the dispersion relation, we proceed numerically due to the fact
that the variable $\zeta $ depends on the energy through the cyclotron
radii. In order to compare with the quantum mechanical solution for $\zeta
\neq 0$ one needs to rewrite Eq.(\ref{esemi}) as a function of \textit{the
center of the classical skipping orbit} which plays the role of $x_{0}$ for
the edge states. For $\zeta\!\sim\! 0$ we have checked that  $%
x_{0}\!\equiv\! x_{+}$ , $x_{0}\!\equiv\! x_{-}$ or $x_{0}\!\equiv%
\!(x_{+}+x_{-})/2$ leads to almost the same dispersion relation. In Fig.\ref
{sce} we plot Eq.(\ref{esemi}) after chossing $x_{0}\equiv (x_{+}+x_{-})/2$. Notice
that the semiclassical solution follows quite satisfactory the quantum
results even in the low energy region and it is almost indistinguishable
from them for $x_{0}\sim 0$.

We end this section by presenting results of the numerical integration of
the semiclassical equations Eq.(\ref{em}) using a fourth order Runge-Kutta
algorithm. The results agree with the analytical solution obtained to 
lowest order in $\hbar $, and they are summarized in Fig.\ref{sce} where the 
skipping orbit with two
radii is clearly observed. The components $n_{1}$ and $n_{2}$ of the
classical vector $\mathbf{n}$ show the expected behavior with a continuous
evolution at the bouncing point. The out of plane component $n_{3}$ is small 
and presents fast changes at the bouncing points.
This component decreases as the energy increases
in agreement with the semiclassical assumption that, to lowest order in 
$\hbar $, predicts  $n_{3}=0$.
\section{Summary and Conclusions}

We have analyzed the eigenstates and the energy spectrum of a 2DEG with
spin-orbit coupling in the presence of a perpendicular magnetic field. We
focused on the edge states that appear when the 2DEG is confined in the
transverse $x$-direction by a square well potential. We first discussed the
low energy states in the high field limit. The rest of our work was devoted
to study the high energy states (high quantum number $n$). In this regime,
the spin-orbit coupling has an important effect on the edge states: while
for the $\alpha \!=\!0$ the edge states with $k=0$ (that corresponds to $%
x_{0}=0$) have an energy separation $\Delta E\!=\!2\hbar \omega _{c}$,\cite
{vanHouten89} for $\alpha \!\neq \!0$ the energy separation is $\hbar \omega
_{c}$, see Eq. (\ref{e0}). As pointed out in section II, the effect of the
spin-orbit coupling increases with $\alpha $ and with the quantum number $n$%
, and there is always a high energy regime where the spin-orbit coupling
dominates. In this high $n$ regime, the energy spectrum of the edge states
follows Eq. (\ref{e0}).

In the bulk, states with large and
small radii are quasi-degenerated. The bouncing at the surface mixes
them leading to hybrid states that combine large 
and small  radii. The
mixing is evident in the quantum solution where the energy spectrum versus $%
x_{0}$ shows avoided level crossings, a fingertip of level mixing.

The spin texture of these states is also discussed in terms of the
classical solution. To lowest order in $\hbar $ the spin lies in the plane
of the 2DEG and its direction is perpendicular to the velocity. The relative
orientation of the spin with respect to the velocity is different along the
segments with large and small cyclotron radius.

The picture obtained with the semiclassical approximation accounts for the
quantum mechanical prediction of a splitted transverse focusing peaks\cite
{UsajB04_focusing} that was recently experimentally observed in hole gas in
GaAs.\cite{RokhinsonGPW04}

Partial financial support by ANPCyT Grant 99 3-6343
and Foundaci\'on Antorchas, Grants 14169/21 and 14116/192, 
are gratefully acknowledged.

\end{document}